\documentclass[10pt]{IEEEtran}

\usepackage{pifont}
\usepackage{amsmath,graphicx}
\usepackage{tabularx}
\usepackage{multirow, xcolor}

\begin{document}
\title{Enhancing Speech Emotion Recognition Through Differentiable Architecture Search}
\author{Thejan Rajapakshe$^{*}$, Rajib Rana,~\IEEEmembership{Member,~IEEE}, Sara Khalifa,~\IEEEmembership{Member,~IEEE}, Berrak Sisman,~\IEEEmembership{Member,~IEEE} and Bj\"{o}rn W.\ Schuller,~\IEEEmembership{Fellow,~IEEE}
    \thanks{Thejan Rajapakshe \& Rajib Rana is with University of Southern Queensland, Australia}
    \thanks{Sara Khalifa is with Queensland University of Technology, Australia}
    \thanks{Berrak Sisman is with Speech \& Machine Learning Lab, The University of Texas at Dallas, USA}
    \thanks{Bj\"{o}rn W.\ Schuller is with GLAM, Imperial College London, UK}
    \thanks{$^{*}$Thejan.Rajapakshe@usq.edu.au}%
}

\maketitle
\IEEEpeerreviewmaketitle

\begin{abstract}
Speech Emotion Recognition (SER) is a critical enabler of emotion-aware communication in human-computer interactions. Recent advancements in Deep Learning (DL) have substantially enhanced the performance of SER models through increased model complexity. However, designing optimal DL architectures requires prior experience and experimental evaluations. Encouragingly, Neural Architecture Search (NAS) offers a promising avenue to determine an optimal DL model automatically. In particular, Differentiable Architecture Search (DARTS) is an efficient method of using NAS to search for optimised models. This paper proposes a DARTS-optimised joint CNN and LSTM architecture, to improve SER performance, where the literature informs the selection of CNN and LSTM coupling to offer improved performance. While DARTS has previously been applied to CNN and LSTM combinations, our approach introduces a novel mechanism, particularly in selecting CNN operations using DARTS. In contrast to previous studies, we refrain from imposing constraints on the order of the layers for the CNN within the DARTS cell; instead, we allow DARTS to determine the optimal layer order autonomously. Experimenting with the IEMOCAP and MSP-IMPROV datasets, we demonstrate that our proposed methodology achieves significantly higher SER accuracy than hand-engineering the CNN-LSTM configuration. It also outperforms the best-reported SER results achieved using DARTS on CNN-LSTM.
\end{abstract}


\section{Introduction}
\IEEEPARstart{R}{ecognising} the emotion embedded in speech is a crucial but challenging problem. Encouragingly, Speech Emotion Recognition (SER) research has made significant progress in the last decade utilising the steep rise of deep learning~\cite{Zhao2019SpeechNetworks, Jalal2020EmpiricalRecognition, Lieskovska2021AMechanism, Latif2022MultitaskRecognition}. Deep learning offers the ability to learn features automatically; however, finding the right deep-learning architecture for SER is challenging as the model needs to be gradually modified and trained recursively until the best configuration is found. This can be prohibitively time-consuming, given the time needed to train and test numerous configurations.

An alternative to the conventional approach is ``neural architecture search" (NAS), which can help discover the optimal neural network for a given task. In NAS, the search is conducted over a discrete set of candidate operations. This requires the model to be trained on a specific configuration before moving on to the next configuration. This is, however, time-consuming~\cite{Ren2021ASearch}. The differentiable architecture search (DARTS)~\cite{Liu2018DARTS:SEARCH} found a way of relaxing the discrete set of candidate operations, allowing the search space to be continuous. Researchers show that DARTS can decrease the computation time of 2000 GPU days of reinforcement learning or 3150 GPU days of evolution to 2–3 GPU days~\cite{Liu2018DARTS:SEARCH, Zoph2017LearningRecognition}. 
This motivates us to focus on DARTS in this paper.

Furthermore, previous studies have shown that a multi-temporal CNN stacked on LSTM offers to capture contextual information at multiple temporal resolutions, complementing LSTM for modelling long-term contextual information, thus offering improved performance~\cite{Han2018TowardsRecognition, Haque2018ImageAttention, Li2019ImprovedLearning, Latif2022SelfRecognition}. Therefore, this paper aims to utilise DARTS for a joint CNN LSTM configuration. 
Figure~\ref{fig:ovearall_architecture_darts} shows the overview of our proposed architecture. 

In the literature, researchers have used DARTS in SER tasks~\cite{Sun2022EmotionNAS:Recognition, Wu2022NEURALRECOGNITION}; however, several manual processes are involved in their methods. For example, the most relevant study to our proposal is \cite{Wu2022NEURALRECOGNITION}, where authors predefined the order of several layers, structures and operations from where DARTS make the optimum selection of parameters for those layers. In contrast, our proposed methodology offers a novel mechanism that minimises the need to predefine the layers, offering improved autonomy. The key contributions of this paper are summarised as follows: \begin{enumerate}
    \item We propose a novel DARTS-optimised joint CNN and LSTM architecture showcasing the feasibility of using network architecture search methods for SER.
    \item Unlike previous research, our approach provides greater autonomy to DARTS in selecting optimal network configurations. 
    \item Experimental results conducted with IEMOCAP and MSP-IMPROV datasets validate that the SER model obtained through our approach surpasses the performance achieved in prior studies. 
\end{enumerate}

\section{Related Work}
In this section, we primarily discuss the application of Neural Architecture Search and DARTS for speech emotion recognition. 
Our findings highlight a shortage of studies, emphasising the necessity for additional exploration of NAS and DARTS in the field of SER.

The initial paper proposing the application of NAS in SER employs a controller network that shapes the architecture by the number of layers and nodes per layer and the hyperparameter activation function of a child network by reinforcement learning~\cite{Zhang18-ELF}. The authors show a competitive improvement over human-designed architectures.

EmotionNAS is a two-branch NAS strategy introduced by Sun H. et al.~\cite{Sun2022EmotionNAS:Recognition} in 2022. The authors use DARTS to optimise the two models in the two branches separately, the CNN model and RNN model, which use a spectrogram and a waveform as inputs, respectively. They obtained an unweighted accuracy of $69.1\%$ from the combined model for the IEMOCAP dataset. They also report the performance of $63.5\%$ in the spectrogram branch, which only uses a CNN based layer architecture.
The key distinction between our approach and EmotionNAS is that we use an LSTM layer coupled in series with the CNN layer as in Figure~\ref{fig:ovearall_architecture_darts}. In contrast, EmotionNAS uses an RNN layer parallel to the CNN layer in a different branch. Moreover, we integrate the attention in LSTM to enhance the overall performance of the CNN-LSTM coupling for SER further.

Wu X. et al.~\cite{Wu2022NEURALRECOGNITION} used SER as their DARTS application and proposed a uniform path dropout strategy to optimise candidate architecture. They used the IEMOCAP dataset to develop an SER model with an accuracy of $56.28\%$ for a four-class classification problem using discrete Fourier transform spectrograms extracted from audio as input. 
Here, authors apply DARTS on CNN and LSTM with attention for SER. This is a preliminary study where authors predefine several configurations of layers, structures and operations and let DARTS select only from those configurations. We aim to extend this study by allowing DARTS to select the best network architecture without providing any predefined configurations.

Liu et al.~\cite{Liu2023SpeechLearning} utilised an attention-based bi-directional LSTM followed by a CNN layer for a SER problem. They have achieved a significant performance of $66.27\%$ for the IEMOCAP Dataset. Their idea of `CNN - LSTM attention' paved the foundation for our model architecture.

\section{DARTS framework for CNN and LSTM coupling}

\begin{figure}[!t]
    \centering
    \includegraphics[width=\linewidth]{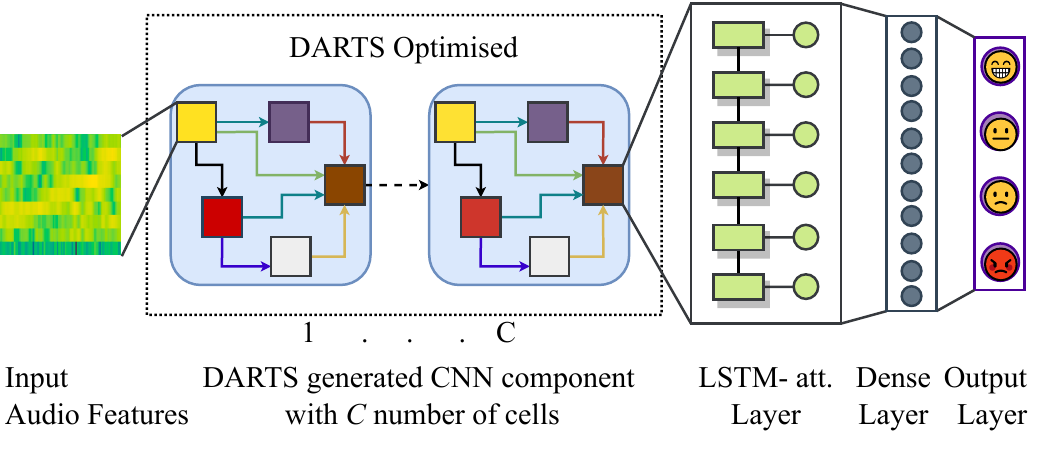} 
    \caption{The proposed model architecture comprises input features processed through CNN, LSTM, and Dense layers, and utilises DARTS for optimising the CNN component.}
    \label{fig:ovearall_architecture_darts}
\end{figure}

The proposed methodology applies DARTS to enhance SER accuracy using a CNN LSTM network, influenced by research demonstrating improved performance when combining CNN and LSTM layers~\cite{Han2018TowardsRecognition, Li2019ImprovedLearning, Latif2022SelfRecognition, Liu2023SpeechLearning}. DARTS is first applied to CNN, which is then fused with LSTM. The LSTM layer follows the DARTS-optimised CNN, facilitating joint training for loss minimisation (Figure ~\ref{fig:ovearall_architecture_darts}).

DARTS uses a differentiable approach to network optimisation. The building block of the DARTS algorithm is a computation cell. It seeks to optimise the cell to gain maximum performance from the architecture. A DARTS cell is modelled as a directed graph where each node is a representation, and the edge is an operation that can be applied to a representation. 

\begin{figure}[!t]
    \centering
    \includegraphics[width=\linewidth]{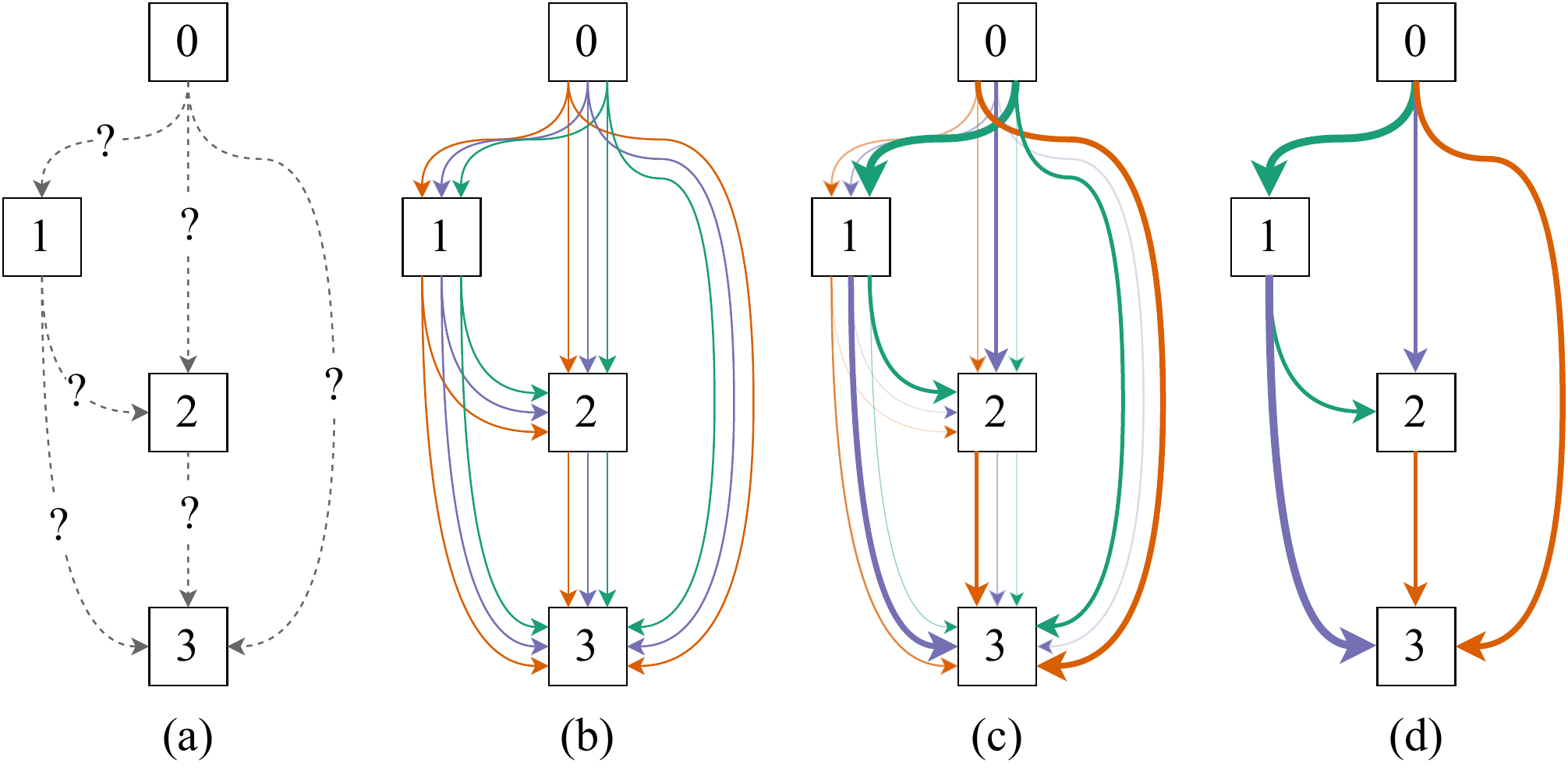}
    \caption{DARTS employs steps (a) to (d) to search cell architectures: (a) initialises the graph, (b) forms a search space, (c) updates edge weights, and (d) determines the final cell structure. Edges represent operations, nodes signify representations, with light-coloured edges indicating weaker and dark-coloured edges representing stronger operations.}
    \label{fig:DARTS_structure}
\end{figure}

To use the DARTS cell in a CNN network, we model the node as a feature map and an edge for an operation. One speciality of this graph is that each node connects by an edge with all its preceding nodes as in Figure~\ref{fig:DARTS_structure} (a). If the output of the node $j$ is $x^{(j)}$ and operation on the edge connecting the nodes $i$ and $j$ is $o^{(i,j)}$, 
$x^{(j)}$ can be obtained by the Equation~\ref{eq:output_from_node}:
\begin{equation}
    \centering
    \label{eq:output_from_node}
    x^{(j)} = \sum_{i<j} o^{(i,j)}(x^{(i)})
\end{equation}

Initially, the candidate search space is created by connecting each node of the DARTS cell with a set of operations as shown in Figure~\ref{fig:DARTS_structure} (b). A weight parameter `$\alpha$' is introduced to Equation~\ref{eq:output_from_node} to find the optimum edge (operation) between two nodes, $i$ and $j$, out of the candidate search space of all the operations. The output from the node can be expressed as in Equation~\ref{eq:output_from_node_weighted}.
\begin{equation}
    \centering
    \label{eq:output_from_node_weighted}
    x^{(j)} = \sum_{i<j} \alpha^{(i,j)}o^{(i,j)}(x^{(i)})
\end{equation}
Then the continuous relaxation of the search space updates the weights ($\alpha^{i,j}$) of the edges. The final architecture can be obtained by selecting the operation between two nodes with the highest weight $(o^{(i,j)*})$ by using Equation~\ref{eq:highest_weight_edge}.
\begin{equation}
    \centering
    \label{eq:highest_weight_edge}
    o^{(i,j)*} = argmax_o(\alpha^{(i,j)})
\end{equation}
Searched discrete cell architecture can be shown in Figure~\ref{fig:DARTS_structure} (d).

The number of cells ($C$) in a model is a 
parameter to the DARTS algorithm which defines how many DARTS cells are stacked to create the model. Each cell uses the output from the last two cells as the input. If output from each cell $t$ is $y_t$ and the function inside the cell is $f$, $y_t$ can be expressed as; 
\begin{equation}
    \centering
    \label{eq:layer_output}
    y_t = f(y_{t-1}, y_{t-2})
\end{equation}

DARTS uses two types of CNN cells, namely `normal' and `reduction' cells. 
It sets the stride as 1 in normal cells and 2 in reduction cells, so the output is down-sampled in reduction cells. This down-sampling enables the model to remove the redundancy of intermediate features and reduce the complexity.

\section{Experimental Setup}

\subsection{Dataset and Feature Selection}
We use the widely used IEMOCAP~\cite{Busso2008IEMOCAP:Database} and MSP-IMPROV \cite{Busso2017MSP-IMPROV:Perception} datasets for our experiments. 
Our study uses the improvised subset of IEMOCAP and the four categorical labels, happiness, sadness, anger, and neutral as classes from the datasets. 
We use five-fold cross-validation with at least one speaker out in our training and evaluations. At each fold, the training dataset is divided into two subsets, `search', and `training', by a $70/30$ fraction. The `search' set is used in architecture search; the `training' set is used in optimising the searched architecture, and the remaining testing dataset is used to infer and obtain the testing performance of the searched and optimised model. This way, we manage to split the dataset into three sets in each cross-validation session. Also, this allows the utterances in each split to be mutually exclusive.

In this paper, we use Mel Frequency Cepstral Coefficients (MFCC) as input features to the model. MFCC has been used as the input feature in many SER studies in the literature~\cite{Davis1980ComparisonSentences, Latif2019DirectSpeech} and has obtained promising results. We extract $128$ MFCCs from each $8$-second audio utterance from the dataset. If the audio utterance length is less than $8$ seconds, we added padding with zeros while the lengthier utterances are truncated. The MFCC extraction from Librosa python library~\cite{McFee2015Librosa:Python} outputs a shape $128\times512$, downsampled with max pooling to create a spectrogram of the shape $128\times128$.

\subsection{Baseline Models}
We compare the performance of our methodology with three hand-engineered baseline models: 1) CNN, 2) CNN+ LSTM, and 3) CNN+LSTM with attention. The CNN baseline model consists of one CNN layer (kernel size=2, stride=2, and padding=2) followed by a Max-Pooling layer (kernel size=2 and stride=2). Two dense layers then consume the output from the Max-Pooling layer after applying a dropout of $0.3$. Finally, the last dense layer has four output units depicting the four emotion classes, and the model outputs the probability estimation of each emotion for a given input by a Softmax function. This model architecture is inspired by the CNN+LSTM SER model by Etienne C. et al.~\cite{Etienne2018CNN+LSTMAugmentation}.

The CNN+LSTM baseline model is built, including an additional bi-directional LSTM layer of 128 units after the Max-Pooling layer. An attention layer is added to the LSTM layer in the `CNN+LSTM attention' baseline model.

\subsection{DARTS Configuration}
The DARTS cell search space consists of pooling operations such as $3\times3$ max pooling and average pooling, convolutional operations such as $3\times3$ and $5\times5$ separable convolutions, $3\times3$ and $5\times5$ dilated convolution, identity connections, and no connections.
The stochastic gradient descent is used with a learning rate using a cosine annealing schedule as the optimiser to optimise the weights of the operations.
The search is run for $300$ epochs. 
In our experiments, we set $C=4$, setting four DARTS cells.
As defined in~\cite{Liu2018DARTS:SEARCH}, we use reduction cells at every \(\frac{1}{3}C^{th}\) and \(\frac{2}{3}C^{th}\) position of the layers. We randomly initialise $\alpha$ values and the DARTS search algorithm optimises $\alpha$ values related to each operation.

\subsection{Model Configuration}
In this study, the output from the CNN component is passed through to the LSTM layer with 256 units as a vector after flattening the 2D matrix from CNN.
Attention is introduced into the LSTM layer by combining the attention module with the output of the LSTM layer.

We use the popular deep learning library PyTorch 
for model development and training. The experiments are run on a NVIDIA A40-24Q GPU with 24GB of VRAM. The implementation code can be found on GitHub Repository\footnote{https://github.com/jayaneetha/NAS-for-SER}





\section{Evaluation}
\begin{figure*}[!t]
    \centering
    \includegraphics[width=\textwidth]{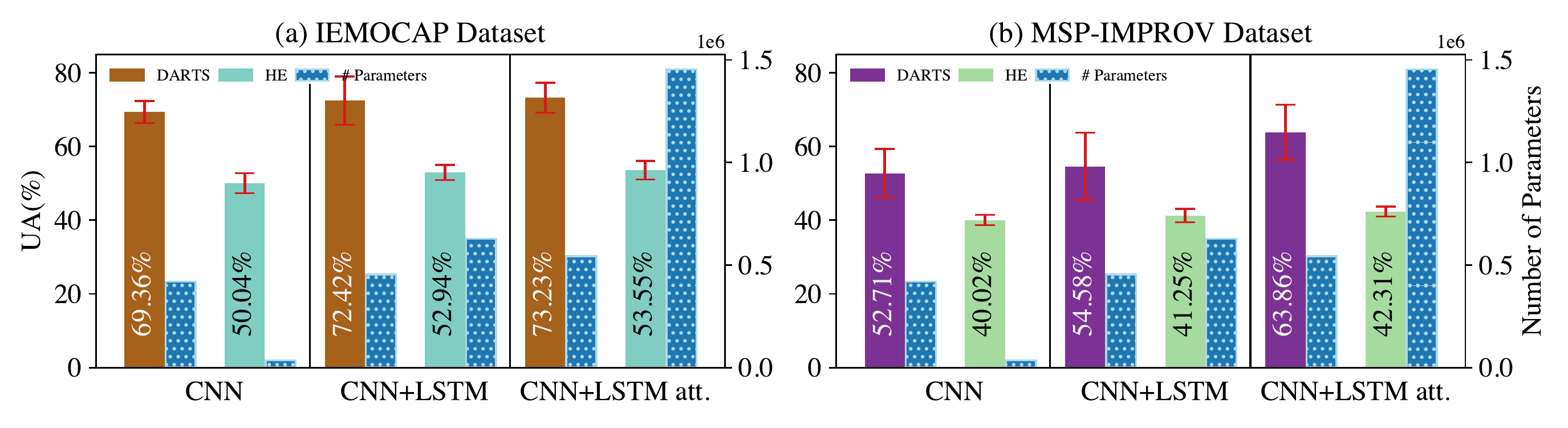}
    \caption{Comparison of UA\% and Number of Parameters between the DARTS generated (DARTS)($C=4$) and Hand Engineered (HE), CNN, CNN+LSTM and CNN+LSTM with attention models for (a) IEMOCAP and (b) MSP-IMPROV datasets.}
    \label{fig:ua_comparison}
    
\end{figure*}

We use Unweighted Accuracy (UA) as obtained by dividing the sum of the recall of all classes by their number, to report our results. This is known to reflect imbalanced data tasks well, as is usual in SER~\cite{Etienne2018CNN+LSTMAugmentation}. 
We also measure the number of parameters of the model by summation of all the trainable parameters in the built model as the representation of the complexity of the model.

First, we compare the performance of the DARTS generated CNN model (CNN -- DARTS) with our benchmark Hand Engineered CNN model (CNN -- HE). The results are presented in Table~\ref{tab:baseline_vs_nas_cnn_only}, showing that the DARTS-generated CNN model outperforms the hand-engineered SER model. This table also shows the performance of the DARTS-generated model with eight cells ($C=8$): it performs poorer than the one with four cells ($C=4$). This is due to the increased complexity of the model; it gets overfitted~\cite{Sun2022EmotionNAS:Recognition}. 

In Table~\ref{tab:baseline_vs_nas_cnn_only}, we also compare the performance of our model with the results of EmotionNAS~\cite{Sun2022EmotionNAS:Recognition}. The authors have implemented NAS in two branches; in one branch, they have CNN, and in the other branch, they have RNN, where the inputs are spectrogram and waveform, respectively. 
We present the combined CNN + RNN performance and performance of the CNN branch in Table~\ref{tab:baseline_vs_nas_cnn_only}. The CNN -- DARTS outperforms both.
It is also notable by comparing the number of parameters and accuracy, the 69.36 UA\% accuracy by CNN -- DARTS was obtained by a model only with 417\,612 parameters whereas the EmotionNAS model contains 2\,370\,000 parameters. This shows that CNN--DARTS model not only outperforms but reduces the size of the model as well. The ease of maintaining the model is enhanced by this. We also report weighted accuracy (WA) in Table~\ref{tab:baseline_vs_nas_cnn_only} as it was reported in the EmotionNAS paper. Similar to using UA, CNN-DARTS outperforms EmotionNAS while using WA.

\begin{table}[!t]
\renewcommand{\arraystretch}{1.3}
\caption{Performance comparison between the DARTS generated CNN model (CNN -- DARTS) and hand-engineered benchmark model (CNN -- HE) for the IEMOCAP dataset.}
    \label{tab:baseline_vs_nas_cnn_only}
    \centering
\begin{tabular}{p{0.215\linewidth}rrcc}
\hline 
\textbf{Model}                      & \multicolumn{1}{l}{\textbf{Param.}} & \multicolumn{1}{l}{\textbf{Cell}}    & \textbf{UA (\%)} & \textbf{WA (\%)}              \\ \hline
CNN -- DARTS                        & \textbf{417\,612}                   & 4                                    & \textbf{69.36 $\pm$ 3.00} & $72.55 \pm 3.70$     \\
CNN -- DARTS                        & 428\,812                            & 8                                    & $62.25 \pm 6.74$    & $63.78 \pm 6.83$          \\
CNN -- HE                           & 35\,017                             & -                                    & $50.04 \pm 2.69$   & $51.01 \pm 2.23$           \\
EmotionNAS [CNN]                    & 130\,000                            & 3                                    & 57.3              & 63.2            \\ 
\small{EmotionNAS [CNN + RNN]}      & 2\,370\,000                                   & 3                                    & 69.1                & 72.1          \\ 
\hline
\end{tabular}
\renewcommand{\arraystretch}{1}
\end{table}

We now compare the performance of the CNN model generated by DARTS (CNN -- DARTS), CNN+LSTM model generated by DARTS (CNN+LSTM -- DARTS), CNN+LSTM with attention model generated by DARTS (CNN+LSTM att. -- DARTS),
and hand-engineered CNN+LSTM (CNN+LSTM -- HE). We perform these comparisons using both IEMOCAP and MSP-IMPROV datasets and present the results in Figure~\ref{fig:ua_comparison} (a) and Figure~\ref{fig:ua_comparison} (b), respectively. We note that for both datasets, (i) `CNN+LSTM att. -- DARTS' performs the best compared to CNN -- DARTS and CNN+LSTM -- DARTS; (ii) DARTS optimised models perform significantly better than the hand-engineered models; (iii) DARTS can produce higher-performing (accuracy) models with fewer parameters than hand-engineered models. 
In addition to the above comparisons, `CNN+LSTM att. -- DARTS' produces better results compared to Wu X. et al.~\cite{Wu2022NEURALRECOGNITION}, who obtained 56.28\% UA for the CNN+RNN attention model for the IEMOCAP dataset. 

\begin{figure}[t]
    \centering
    \includegraphics[width=\linewidth]{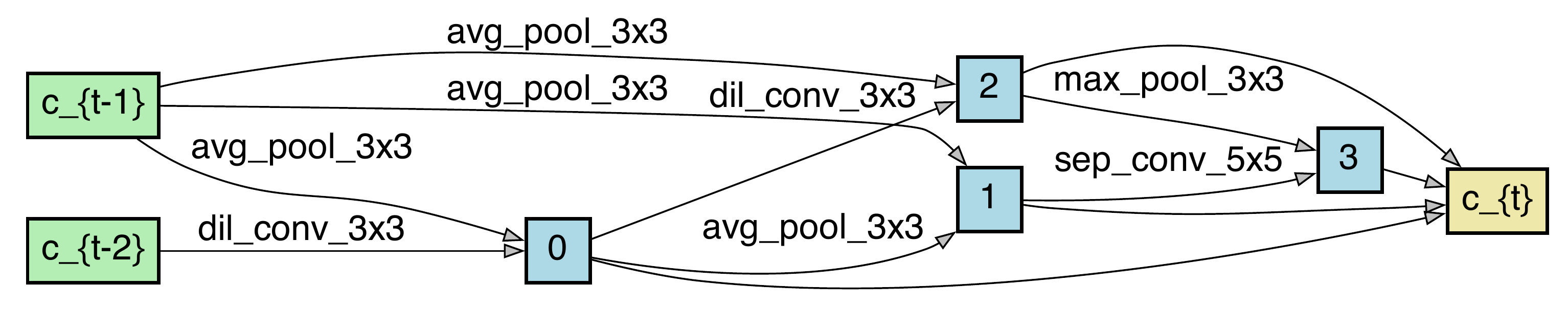}
    \includegraphics[width=\linewidth]{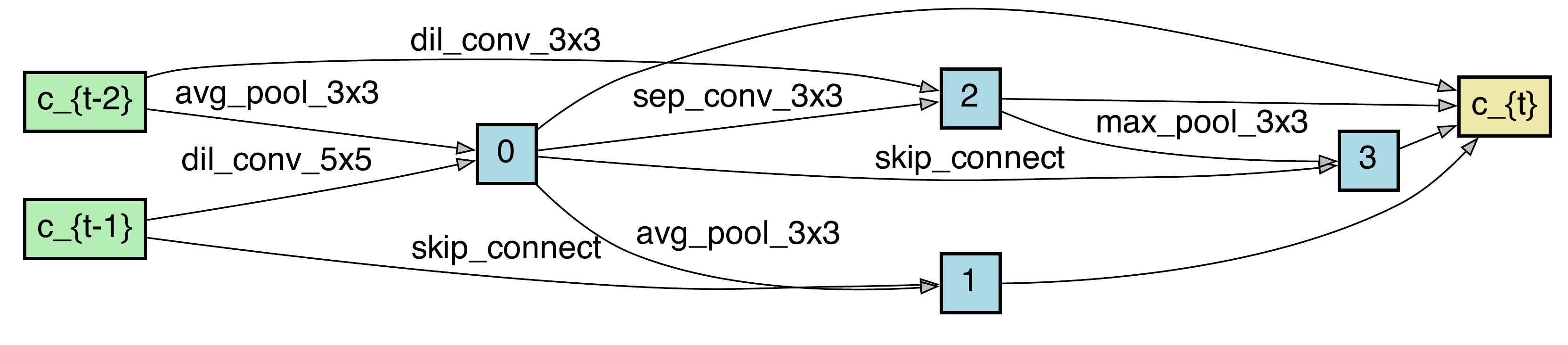}
    \caption{DARTS searched $t^{th}$ cell structure for Normal Cell (Top) Reduction Cell (Bottom) for the CNN+LSTM att. -- DARTS  model.}
    \label{fig:cell_viz}
\end{figure}

By comparing the number of parameters in the Hand Engineered model and DARTS-generated models, it is seen that the Hand Engineered model has a significantly smaller number of parameters. This is because it only contains 1 Convolution 2D layer and 1 Max Pooling layer. However, the DARTS cell has at least four layers producing more parameters. But, as supported by the literature, we conjecture that the improved performance of the proposed `CNN+LSTM att. -- DARTS' model is not due to an increased number of parameters~\cite{Zhang2021UnderstandingGeneralization}. The performance improvement is achieved by allowing DARTS for optimal network selection. For example, Figure~\ref{fig:cell_viz} shows the searched normal and reduction $t^{th}$ cell (c\_\{t\}) structures for the CNN+LSTM att. -- DARTS model generated by DARTS. Note that normal Cell uses pooling layers as the initial set of layers for optimal performance, which deviates from the usual practice of choosing the convolution layer first.

\section{Conclusions}
This study aimed to assess the viability of using the DARTS algorithm to optimise a neural architecture for a joint CNN and LSTM-based SER model. The approach involved augmenting a DARTS-optimised CNN with an LSTM component and jointly training the model to minimise the SER loss. The research findings demonstrate the effectiveness of DARTS in optimising neural architectures, surpassing hand-engineered models. Notably, the study also reveals that as model complexity increases, SER performance decreases, highlighting the superiority of simplified models and emphasising the continued relevance of DARTS in architectural refinement/simplifications for SER models. This research contributes valuable insights into the optimisation of neural architectures for SER applications.

\bibliographystyle{IEEEbib}
\bibliography{references}

\begin{thebibliography}{10}

\bibitem{Zhao2019SpeechNetworks}
Jianfeng Zhao, Xia Mao, and Lijiang Chen,
\newblock ``{Speech emotion recognition using deep 1D {\&} 2D CNN LSTM
  networks},''
\newblock {\em Biomedical Signal Processing and Control}, vol. 47, pp.
  312--323, 1 2019.

\bibitem{Jalal2020EmpiricalRecognition}
Md~Asif Jalal, Rosanna Milner, and Thomas Hain,
\newblock ``{Empirical interpretation of speech emotion perception with
  attention based model for speech emotion recognition},''
\newblock {\em Proceedings of the Annual Conference of the International Speech
  Communication Association, INTERSPEECH}, vol. 2020-October, pp. 4113--4117,
  2020.

\bibitem{Lieskovska2021AMechanism}
Eva Lieskovsk{\'{a}}, Maroš Jakubec, Roman Jarina, Michal Chmul{\'{i}}k,
  Yuan-Fu Liao, Patrick Bours, and Chiman Kwan,
\newblock ``{A Review on Speech Emotion Recognition Using Deep Learning and
  Attention Mechanism},''
\newblock {\em Electronics 2021, Vol. 10, Page 1163}, vol. 10, no. 10, pp.
  1163, 5 2021.

\bibitem{Latif2022MultitaskRecognition}
Siddique Latif, Rajib Rana, Sara Khalifa, Raja Jurdak, and Bjorn~W. Schuller,
\newblock ``{Multitask Learning From Augmented Auxiliary Data for Improving
  Speech Emotion Recognition},''
\newblock {\em IEEE Transactions on Affective Computing}, pp. 1--13, 7 2022.

\bibitem{Ren2021ASearch}
Pengzhen Ren, Yun Xiao, Xiaojun Chang, Po~Yao Huang, Zhihui Li, Xiaojiang Chen,
  and Xin Wang,
\newblock ``{A Comprehensive Survey of Neural Architecture Search},''
\newblock {\em ACM Computing Surveys (CSUR)}, vol. 54, no. 4, pp. 76, 5 2021.

\bibitem{Liu2018DARTS:SEARCH}
Hanxiao Liu, Karen Simonyan, and Yiming Yang,
\newblock ``{DARTS: Differentiable Architecture Search},''
\newblock {\em 7th International Conference on Learning Representations, ICLR
  2019}, 6 2018.

\bibitem{Zoph2017LearningRecognition}
Barret Zoph, Vijay Vasudevan, Jonathon Shlens, and Quoc~V. Le,
\newblock ``{Learning Transferable Architectures for Scalable Image
  Recognition},''
\newblock {\em Proceedings of the IEEE Computer Society Conference on Computer
  Vision and Pattern Recognition}, pp. 8697--8710, 7 2017.

\bibitem{Han2018TowardsRecognition}
Wenjing Han, Huabin Ruan, Xiaomin Chen, Zhixiang Wang, Haifeng Li, and Björn
  Schuller,
\newblock ``{Towards temporal modelling of categorical speech emotion
  recognition},''
\newblock {\em Proceedings of the Annual Conference of the International Speech
  Communication Association, INTERSPEECH}, vol. 2018-September, pp. 932--936,
  2018.

\bibitem{Haque2018ImageAttention}
Kazi~Nazmul Haque, Mohammad~Abu Yousuf, and Rajib Rana,
\newblock ``{Image denoising and restoration with CNN-LSTM Encoder Decoder with
  Direct Attention},''
\newblock {\em arXiv Prepr.}, pp. 1--12, 1 2018.

\bibitem{Li2019ImprovedLearning}
Yuanchao Li, Tianyu Zhao, and Tatsuya Kawahara,
\newblock ``{Improved end-to-end speech emotion recognition using self
  attention mechanism and multitask learning},''
\newblock {\em Proceedings of the Annual Conference of the International Speech
  Communication Association, INTERSPEECH}, vol. 2019-September, pp. 2803--2807,
  2019.

\bibitem{Latif2022SelfRecognition}
Siddique Latif, Rajib Rana, Sara Khalifa, Raja Jurdak, and Bjorn~Wolfgang
  Schuller,
\newblock ``{Self Supervised Adversarial Domain Adaptation for Cross-Corpus and
  Cross-Language Speech Emotion Recognition},''
\newblock {\em IEEE Transactions on Affective Computing}, 2022.

\bibitem{Sun2022EmotionNAS:Recognition}
Haiyang Sun, Zheng Lian, Bin Liu, Ying Li, Licai Sun, Cong Cai, Jianhua Tao,
  Meng Wang, and Yuan Cheng,
\newblock ``{EmotionNAS: Two-stream Neural Architecture Search for Speech
  Emotion Recognition},''
\newblock {\em arXiv Prepr.}, pp. 1--5, 3 2022.

\bibitem{Wu2022NEURALRECOGNITION}
Xixin Wu, Shoukang Hu, Zhiyong Wu, Xunying Liu, and Helen Meng,
\newblock ``{NEURAL ARCHITECTURE SEARCH FOR SPEECH EMOTION RECOGNITION},''
\newblock {\em ICASSP, IEEE International Conference on Acoustics, Speech and
  Signal Processing - Proceedings}, vol. 2022-May, pp. 6902--6906, 2022.

\bibitem{Zhang18-ELF}
Zixing Zhang, Jing Han, Kun Qian, and Björn Schuller,
\newblock ``{Evolving Learning for Analysing Mood-Related Infant
  Vocalisation},''
\newblock {\em Proceedings INTERSPEECH 2018, 19. Annual Conference of the
  International Speech Communication Association}, pp. 142--146, 2018.

\bibitem{Liu2023SpeechLearning}
Zhen~Tao Liu, Meng~Ting Han, Bao~Han Wu, and Abdul Rehman,
\newblock ``{Speech emotion recognition based on convolutional neural network
  with attention-based bidirectional long short-term memory network and
  multi-task learning},''
\newblock {\em Applied Acoustics}, vol. 202, pp. 109178, 1 2023.

\bibitem{Busso2008IEMOCAP:Database}
Carlos Busso, Murtaza Bulut, Chi-Chun Lee, Abe Kazemzadeh, Emily Mower, Samuel
  Kim, Jeannette~N Chang, Sungbok Lee, and Shrikanth~S Narayanan,
\newblock ``{IEMOCAP: interactive emotional dyadic motion capture database},''
\newblock {\em Language Resources and Evaluation}, vol. 42, no. 4, pp. 335,
  2008.

\bibitem{Busso2017MSP-IMPROV:Perception}
C~Busso, S~Parthasarathy, A~Burmania, M~AbdelWahab, N~Sadoughi, and E~M
  Provost,
\newblock ``{MSP-IMPROV: An Acted Corpus of Dyadic Interactions to Study
  Emotion Perception},''
\newblock {\em IEEE Transactions on Affective Computing}, vol. 8, no. 1, pp.
  67--80, 2017.

\bibitem{Davis1980ComparisonSentences}
Steven Davis and Paul Mermelstein,
\newblock ``{Comparison of parametric representations for monosyllabic word
  recognition in continuously spoken sentences},''
\newblock {\em IEEE transactions on acoustics, speech, and signal processing},
  vol. 28, no. 4, pp. 357--366, 1980.

\bibitem{Latif2019DirectSpeech}
Siddique Latif, Rajib Rana, Sara Khalifa, Raja Jurdak, and Julien Epps,
\newblock ``{Direct Modelling of Speech Emotion from Raw Speech},''
\newblock in {\em Proceedings of the Annual Conference of the International
  Speech Communication Association, INTERSPEECH}, 2019, pp. 3920--3924.

\bibitem{McFee2015Librosa:Python}
Brian McFee, Colin Raffel, Dawen Liang, Daniel P~W Ellis, Matt McVicar, Eric
  Battenberg, and Oriol Nieto,
\newblock ``{librosa: Audio and music signal analysis in python},''
\newblock 2015, vol.~8.

\bibitem{Etienne2018CNN+LSTMAugmentation}
Caroline Etienne, Guillaume Fidanza, Andrei Petrovskii, Laurence Devillers, and
  Benoit Schmauch,
\newblock ``{CNN+LSTM Architecture for Speech Emotion Recognition with Data
  Augmentation},''
\newblock in {\em Workshop on Speech, Music and Mind (SMM 2018)}, ISCA, 9 2018,
  ISCA.

\bibitem{Zhang2021UnderstandingGeneralization}
Chiyuan Zhang, Samy Bengio, Moritz Hardt, Benjamin Recht, and Oriol Vinyals,
\newblock ``{Understanding deep learning (still) requires rethinking
  generalization},''
\newblock {\em Communications of the ACM}, vol. 64, no. 3, pp. 107--115, 2
  2021.

\end{thebibliography}

\end{document}